\documentclass[aps,twocolumn,superscriptaddress,footinbib,longbibliography,10pt,preprintnumbers]{revtex4-1}
\usepackage{graphicx}
\usepackage{amsmath}
\usepackage{amssymb}
\usepackage{bm}
\usepackage{mathtools}
\usepackage{hyperref}
\usepackage{tikz}
\usetikzlibrary{calc}
\usepackage{tikz-cd}
\hypersetup{
    linktocpage=true,
    colorlinks=true,
    citecolor=blue,
    linkcolor=blue,
    urlcolor=blue,
}

\DeclareMathOperator{\U}{U}

\DeclareMathOperator{\SU}{SU}

\DeclareMathOperator{\TY}{TY}

\DeclareMathOperator{\Vect}{Vect}

\DeclareMathOperator{\TQFT}{TQFT}

\newcommand{\Sfaith}{\mathcal{S}^{\text{faith}}}
\newcommand{\ssb}{\rightsquigarrow} %

\usepackage[status=draft,inline,nomargin]{fixme}
\fxusetheme{color}
\FXRegisterAuthor{ko}{ako}{KO}

\begin{document}

\preprint{ YITP-26-149}
\preprint{ RIKEN-iTHEMS-Report-26}

\title{Enforced Gaplessness by Combining Non-Anomalous Symmetries}

\author{Takamasa Ando}
\affiliation{Center for Gravitational Physics and Quantum Information, Yukawa Institute for Theoretical Physics, Kyoto University, Kyoto 606-8502, Japan}

\author{Kantaro Ohmori}
\affiliation{RIKEN Center for Interdisciplinary Theoretical and Mathematical Sciences (iTHEMS), RIKEN, Wako, Saitama 351-0198, Japan}

\begin{abstract}
Symmetry-enforced gaplessness is a phenomenon in which a quantum system is forced to be gapless due to its global symmetry.
Previous works have mainly focused on anomalous symmetries.
Another approach, using symmetry spans, which specify two embeddings of a common symmetry into two larger symmetries, was proposed by the authors.
In this work, we show that a symmetry span can enforce gaplessness even when both larger symmetries are invertible and non-anomalous: each alone admits a trivially gapped phase, but because the embeddings differ by a twist associated with a symmetry-protected topological phase, the symmetry they generate admits no gapped phase with finitely many ground states.
We present such a span of a finite group and $\U(1)^2$ in 1+1 dimensions, with continuum realizations and a lattice Hamiltonian, with a finite-dimensional local Hilbert space and finitely many parameters, that is gapless in any region of its parameter space.
We also discuss a 2+1-dimensional example and spans of only finite symmetries.
\end{abstract}

\maketitle

\textit{Introduction}-----
Physical models in nature often show their nontrivial non-perturbative dynamics in their ground states, and usual methods such as perturbation analysis by renormalization group flow or mean-field analysis do not work well in such situations.
Therefore it is important to develop a way of understanding the non-perturbative dynamics by using ultraviolet data of the system.

Global symmetry is one of characteristic properties of a physical system, and can constrain the non-perturbative dynamics of the system.
For instance, an anomaly of a global symmetry constrains the infrared (IR) behavior through anomaly matching \cite{tHooft:1979rat}, requiring the system to be gapless, have degenerate ground states, or exhibit nontrivial topological order.
Constraints on IR phases by anomaly matching for various types of symmetries, including higher-form or non-invertible symmetries, have been studied both in high-energy physics and condensed matter physics \cite{Wang:2014lca, Gaiotto:2014kfa, Wang:2016gqj, Sodemann:2016mib, Gaiotto:2017yup, Wang:2017txt, Chang:2018iay, Cordova:2019bsd, Thorngren:2019iar, Komargodski:2020mxz, Hayashi:2022fkw, Apte:2022xtu, Zhang:2023wlu, Choi:2023xjw, Cordova:2023bja, Antinucci:2023ezl, Brennan:2023ynm, Putrov:2024uor, Antinucci:2025fjp, Lieb:1961fr, koma2000spectral, Oshikawa:2000zwq, Hastings:2003zx, Hastings_2005, roy2012space, Furuya:2015coa, Watanabe:filling, Cheng:2015kce, Cho:2017fgz, Qi2017ground, Huang2017building, Cheng:2018lti, Kobayashi:2018yuk, Levin:2019ifu, Else:2019lft, Ogata:2020hry, Aksoy:2021uxb, Cheng:2022sgb, Seiberg:2023cdc, Yao:2023bnj, Aksoy:2023hve, Seifnashri:2023dpa, Kapustin:2024rrm, Seiberg:2024gek, Liu:2024nox, Ma:2024sig}.

Particular interest of IR scenario in this paper is enforced gaplessness, which is a situation where the system is necessarily gapless in the IR due to the global symmetry.
Previous works on enforced gaplessness utilize anomalous continuous group symmetries and their lattice realizations \cite{Cordova:2019jqi, Thorngren:2020aph, Chatterjee:2024gje, Pace:2024oys, Hsin:2025ria, Pace:2025rfu, Kim:2025tzc, Lew-Smith:2026jqb}.
Another approach by symmetry span is proposed by the authors in \cite{Ando:2026ffy}.
In both cases, enforced gaplessness relies on the existence of either an anomalous invertible symmetry (finite or continuous) or a non-invertible symmetry.
Furthermore, most mechanisms of enforced gaplessness basically rely on continuous group symmetries, whose spontaneous symmetry breaking (SSB) leads to Nambu--Goldstone (NG) gapless modes \cite{Nambu:1960tm, Goldstone:1961eq, Goldstone:1962es}.

In this paper, we provide some classes of simple examples of enforced gaplessness from symmetry spans.
In particular, we provide two classes of symmetry spans that enforce gaplessness:
\begin{enumerate}
  \item Symmetry span by only non-anomalous invertible symmetries.
  \item Symmetry span by only finite symmetries; all of symmetry operators are built from finite numbers of symmetry operators.
\end{enumerate}
The former class is understood almost by using the fundamental knowledge of group theory, and is the main subject of this letter.
The latter class of the span generates non-invertible symmetry with infinitely many simple objects which may have not been studied in the literature.
In both classes, what we can claim from symmetry spans is the absence of any gapped phase with finitely many ground states whose IR limit is described by a relativistic topological quantum field theory (TQFT) \footnote{Thus precisely speaking, one cannot exclude the existence of gapped phases without a TQFT description, such as fracton phases, especially in dimensions higher than two.
Nonetheless, in this paper we somewhat sloppily refer to the absence of TQFT phases as ``gaplessness''.}.
We also provide examples of continuum field theories realizing both classes of symmetry spans, and a lattice realization of the former class of symmetry spans.
The lattice model has a finite-dimensional local Hilbert space and a local Hamiltonian with finitely many parameters.
The model at a particular fine-tuned point was studied in \cite{Vernier:2018han} and known to be gapless. 
The symmetry span argument provides strong support for the claim that the ground states are gapless in any region of the parameter space.

\bigskip
\textit{Enforced gaplessness by symmetry spans}-----
Here we briefly review the general mechanism of enforced gaplessness by symmetry spans \cite{Ando:2026ffy}.
Let $\mathcal{C}$, $\mathcal{D}$, and $\mathcal{E}$ denote categories of symmetries, which may be some higher tensor categories.
For usual symmetries, one can think of them just as groups.
An embedding of symmetry $\mathcal{E}$ into $\mathcal{C}$ is physically realized by a situation where symmetry operators of $\mathcal{E}$ are identified with parts of symmetry operators of $\mathcal{C}$.
Mathematically, this embedding is described by a functor $i_{\mathcal{C}}:\mathcal{E}\to\mathcal{C}$, which is fully faithful and injective on objects.
Then a symmetry span consists of two embeddings of a common symmetry $\mathcal E$ into two symmetries $\mathcal C$ and $\mathcal D$:
\begin{equation}
    \begin{tikzcd}
       \mathcal{E}\arrow[r, hook, "i_{\mathcal{C}}"] \arrow[d, hook', "i_{\mathcal{D}}"'] & \mathcal{C} \arrow[d, hook'] \\
      \mathcal{D} \arrow[r, hook] & \mathcal{S}^{\text{faith}} .
    \end{tikzcd}
\end{equation}
Here, $\Sfaith$ is the full symmetry of the system generated by $\mathcal{C}$ and $\mathcal{D}$. In particular, it contains both $\mathcal{C}$ and $\mathcal{D}$ as sub-categories.

To state the gaplessness criterion, let $\TQFT(\mathcal{E})$ denote the category of gapped theories (TQFTs) with $\mathcal{E}$ symmetry, and same for $\mathcal{C}$ and $\mathcal{D}$.
Roughly, an element of $\TQFT(\mathcal{E})$ consists of all compatible symmetry actions of $\mathcal{E}$ on collections of gapped ground states.
Via the embedding functors $i_{\mathcal{C}}$ and $i_{\mathcal{D}}$, every gapped theory with $\mathcal{C}$ or $\mathcal{D}$ symmetry restricts to a gapped theory with $\mathcal{E}$ symmetry, which allows us to define sub-categories of $\TQFT(\mathcal{E})$: the pullbacks $i_{\mathcal{C}}^*\TQFT(\mathcal{C})$ and $i_{\mathcal{D}}^*\TQFT(\mathcal{D})$.
The definition of pullback categories is physically clear: just restrict the symmetry action of $\mathcal{C}$ or $\mathcal{D}$ to the sub-symmetry $\mathcal{E}$.
Since any gapped phase with the full symmetry $\Sfaith$ must simultaneously be a valid $\mathcal{C}$-symmetric phase and a valid $\mathcal{D}$-symmetric phase, it must lie in both pullback categories.
Thus, a necessary condition for gapped phases to exist is:
\begin{equation}\label{eq:gapped condition}
    i_{\mathcal{C}}^*\TQFT(\mathcal{C}) \cap i_{\mathcal{D}}^*\TQFT(\mathcal{D}) \neq \{0\}.
\end{equation}
When the intersection is trivial, no gapped theory can realize both $\mathcal{C}$ and $\mathcal{D}$ symmetries, and the system is necessarily gapless---this is symmetry enforced gaplessness from a symmetry span.

We emphasize that \eqref{eq:gapped condition} can be violated even when none of $\mathcal{C}$, $\mathcal{D}$, and $\mathcal{E}$ is anomalous.
In such a case, non-anomalous pieces generate a total symmetry $\Sfaith$ that is as anomalous as a continuous symmetry with a perturbative anomaly: it admits no gapped phase with finitely many ground states, whether symmetric or spontaneously broken.
Mathematically, $\Sfaith$ admits no module category with finitely many simple objects.
In particular, $\Sfaith$ cannot be a fusion category, since its regular module category would give such a gapped phase \cite{Huang:2021zvu}. Note that $\Sfaith$ is in general not finite; it can be infinite, or even continuous and infinite-dimensional, and we lack a satisfactory framework to describe such symmetries and their anomalies directly.
The advantage of symmetry spans is that \eqref{eq:gapped condition} involves only the embeddings $i_{\mathcal{C}}$ and $i_{\mathcal{D}}$, so enforced gaplessness can be established without knowing $\Sfaith$ itself, let alone formulating its anomaly.

\bigskip
\textit{Enforced gaplessness by non-anomalous invertible symmetry}-----
Consider a symmetry span of the form
\begin{equation}
  K\hookleftarrow H\hookrightarrow G,
\end{equation}
where $H$ and $G$ are finite groups and $K$ is a compact continuous group.
Here we ask whether such a span can enforce gaplessness when all three symmetries are non-anomalous.
Since $G$ and $K$ are non-anomalous, each admits a trivial symmetry-protected topological (SPT) phase. Under an ordinary subgroup restriction, both remain trivial as $H$-SPT phases, so these phases appear to be compatible.
However, an embedding of symmetries contains more information than an embedding of groups.
In spacetime, symmetry operators can meet at junctions, and an embedding must specify how these junctions are identified as well as the operators themselves. This additional data allows an SPT twist, whose effect on a gapped phase is equivalent to stacking an $H$-SPT phase after the ordinary subgroup restriction \cite{Ando:2026ffy}.
Thus, in $d+1$ spacetime dimensions, a general embedding from $H$ symmetry to $G$ symmetry is specified by a pair $(\phi,\beta)$, where $\phi:H\hookrightarrow G$ is an embedding homomorphism of groups and $\beta\in\mathrm{H}^{d+1} (H,\U(1))$ is a cocycle specifying the SPT twist.
This embedding induces a pullback map on the SPT phases:
\begin{equation}
  \begin{gathered}
    i_{(\phi,\beta)}^*\colon \mathrm{H}^{d+1}(G,\U(1))\to \mathrm{H}^{d+1}(H,\U(1)),\\
    i_{(\phi,\beta)}^*(\alpha)=\phi^*\alpha\cdot\beta,
  \end{gathered}
\end{equation}
which is not a group homomorphism but an affine map.
Here and below, $\U(1)$-valued cochains are written multiplicatively, while classes in finite cohomology groups such as $\mathrm{H}^2(\mathbb{Z}_3^2,\U(1))\cong\mathbb{Z}_3$ are labeled additively.

For groups, without the twist $\Sfaith$ would naively be (a quotient of) the amalgamated product $G*_HK$; heuristically, the twist $\beta$ instead equips $G*_HK$ with some anomaly, which is invisible on $G$ and on $K$.

Suppose that $K$ is connected and admits only the trivial SPT phase, and that its embedding is untwisted, so its pullback is the trivial $H$-SPT phase.
We choose the relative SPT twist $\beta$ so that no $G$-SPT phase pulls back to this phase.
In 1+1D, any compatible gapped phase must therefore spontaneously break at least one of the two symmetries.
However, any gapped phase must preserve $K$, and hence $H$, since breaking $K$ would produce NG modes.
Thus, the finite group $G$ has to be spontaneously broken to a proper subgroup containing $H$.
To see which SSB patterns are compatible with the symmetry span, we need the general classification of SSB patterns together with the SPT phases carried by the unbroken symmetries.

We leave the general classification to Appendix~\ref{app:finite-group-classification} of the End Matter, and here we illustrate it with an example of non-anomalous group symmetry $G=\mathbb{Z}_3\times S_3$ in 1+1D.
A general gapped phase is a finite direct sum of indecomposable ones, in which $G$ permutes the ground states transitively, so it suffices to classify the latter.
First, the SSB patterns $G\ssb L$ of indecomposable phases are classified by subgroups $L\subseteq G$ up to conjugation.
Let us write $S_3=\langle r,s\mid r^3=s^2=1,rs=sr^{-1}\rangle$ and $\mathbb{Z}_3=\langle t\mid t^3=1\rangle$.
For this group, representatives of the nine conjugacy classes of subgroups are
\begin{equation}
  \begin{gathered}
    L=1,\quad \langle s\rangle\cong\mathbb{Z}_2,\quad \langle r\rangle\cong\mathbb{Z}_3,\quad \langle t\rangle\cong\mathbb{Z}_3,\\
    \langle rt\rangle\cong\mathbb{Z}_3,\quad \langle r,s\rangle\cong S_3,\quad
    \langle s\rangle\times\langle t\rangle\cong\mathbb{Z}_2\times\mathbb{Z}_3,\\
    \langle r\rangle\times\langle t\rangle\cong\mathbb{Z}_3^2,\quad G.
  \end{gathered}
\end{equation}
The choice $L=G$ corresponds to no SSB.
All choices except $L=\langle r\rangle\times\langle t\rangle$ have $\mathrm{H}^2(L,\U(1))=0$ and hence admit only the trivial SPT phase.
In the following, we consider the case where $L=\langle r\rangle\times\langle t\rangle\cong\mathbb{Z}_3^2$.
The two ground states of the SSB phase $G\ssb L$ are labeled by
\begin{equation}
  G/L=\{L,sL\}\cong\mathbb{Z}_2.
\end{equation}
The generators $r$ and $t$ preserve each ground state sector, while $s$ exchanges the two sectors.
We note that the unbroken symmetry in the sector $sL$ is actually $sLs^{-1}$ whereas it is isomorphic to $L$.

For each ground state sector, the unbroken symmetry $L$ can carry an SPT phase labeled by an element of $\mathrm{H}^2(L,\U(1))\cong\mathbb{Z}_3$.
However, the choice of SPT phase in the two sectors is not independent, because conjugation by $s$ exchanges the two sectors and acts on the SPT two-cocycles.
Specifically, let $c_{s}(g)=sgs^{-1}\,(g\in G)$ and $\omega\in\mathrm{H}^2(L,\U(1))$ be a two-cocycle representing an SPT phase in the $L$ sector.
Then the SPT phase in the $sL$ sector is represented by the two-cocycle $c_s^*\omega$, which is explicitly given by
\begin{equation}
  c_s^*\omega(g_1,g_2)=\omega(sg_1s^{-1},sg_2s^{-1})\quad(g_1,g_2\in L).
\end{equation}
Then one can check that $c_s^*$ acts on $k\in\mathrm{H}^2(L,\U(1))\cong\mathbb{Z}_3$ by $k\mapsto -k$.
Therefore, classification of SPT degrees over these two ground states is either $(0,0)$ or $(1,2)$.

\medskip
Using the classification above, we can find a symmetry span that enforces gaplessness with only non-anomalous invertible symmetries in 1+1D, which is given by the following symmetry span:
\begin{equation}\label{eq:span_invertible_S3}
  \U(1)^2\hookleftarrow\mathbb{Z}_3^2\xhookrightarrow{\beta}\mathbb{Z}_3\times S_3,
\end{equation}
where $\beta\in\mathrm{H}^2(\mathbb{Z}_3^2,\U(1))$ is a nontrivial cocycle.
To see that this span enforces gaplessness, we first note that preserving all symmetries is not allowed due to the nontrivial SPT twist $\beta$.
Further, the left embedding $\mathbb{Z}_3^2\hookrightarrow\U(1)^2$ enforces the $\mathbb{Z}_3^2$ symmetry to be unbroken, if gapped.
Suppose that the right embedding shows an SSB pattern $G\ssb L$ with $L=\mathbb{Z}_3^2$.
Then the pullback of the trivial $\U(1)^2$ SPT phase is the trivial $\mathbb{Z}_3^2$ SPT phase.
On the other hand, the SPT phases over the unbroken symmetry $L$ coming from the right embedding are either both nontrivial or two inequivalent SPT phases, as explained above.
Hence, the symmetry span \eqref{eq:span_invertible_S3} enforces gaplessness.

\medskip
Let us give a continuum realization of this symmetry span using free compact bosons $\phi_i\sim\phi_i+2\pi$.
We denote their integer momentum and winding charges by $P_i$ and $W_i$, respectively.
For \eqref{eq:span_invertible_S3}, consider two decoupled compact bosons, with $c=2$, and take the $\U(1)^2$ charges to be
\begin{equation}
  Q_1=P_1+3W_2,\qquad Q_2=P_2-3W_1.
\end{equation}
This continuous $\U(1)^2$ symmetry is non-anomalous because the mixed anomalies between the shift and winding symmetries cancel each other.
The shifts $e^{2\pi iP_i/3}$ and the reflection $\phi_2\mapsto-\phi_2$ generate $\mathbb{Z}_3\times S_3$.
Since $e^{2\pi iQ_i/3}=e^{2\pi iP_i/3}$, the two symmetries share the same $\mathbb{Z}_3^2$ subgroup.
The relative SPT class of the two embeddings is a generator $\beta$ of $\mathrm{H}^2(\mathbb{Z}_3^2,\U(1))$ \footnote{This can be checked by restricting the background-field couplings of the two symmetry groups to the common $\mathbb{Z}_3^2$ subgroup and computing the ratio of the resulting partition functions.}.
In this realization, the reflection $\phi_2\mapsto-\phi_2$ maps the $\U(1)^2$ symmetry to a different $\U(1)^2$ with charges $P_1-3W_2$ and $-P_2-3W_1$, and together they generate the full momentum and winding symmetry $\U(1)^4$, whose mixed anomaly by itself forbids a gapped phase.
This is a concrete instance of the anomalous total symmetry $\Sfaith$ discussed above.
The span argument does not rely on it, and applies equally to the lattice model below, where the symmetries are expected to generate an infinite-dimensional, Onsager-like symmetry \cite{Vernier:2018han} that reduces to an anomalous finite-dimensional Lie group only in the IR.

We provide another example of a similar symmetry span enforcing gaplessness and its continuum realization in Supplemental Material, Sec.~\ref{sm:subgroup-data}.
We also discuss symmetry spans in higher dimensions in Appendix~\ref{app:higher-dimensional} of the End Matter.

\bigskip

\textit{Lattice Model}-----
Here we provide a lattice realization of the symmetry span \eqref{eq:span_invertible_S3} and a local gapless Hamiltonian with the symmetry span.
Specifically, we realize an on-site $\mathrm{U}(1)^2$ symmetry and implement the relative SPT twist by conjugating an on-site $\mathbb{Z}_3\times S_3$ action with an SPT disentangler.
Consider a one-dimensional quantum spin chain with a local Hilbert space $\mathcal{H}_j$ at each site $j$. Here, we take the local Hilbert space as
$\mathcal{H}_{j}\cong \mathbb{C}^{|\mathbb{Z}_3|}\otimes\mathbb{C}^{|S_3|}\cong \mathbb{C}_A^3\otimes\mathbb{C}_B^3\otimes\mathbb{C}^2$.
To define symmetry operators, we write the span as $\mathbb{Z}_3^A\times S_3 \hookleftarrow \mathbb{Z}_3^A\times\mathbb{Z}_3^B \hookrightarrow \U(1)^2$.
The $\mathbb{Z}_3^A\times\mathbb{Z}_3^B$ symmetry is generated by
\begin{equation}
  U_A\coloneqq \prod_{j}X_{j,A},\quad U_B\coloneqq \prod_{j}X_{j,B}.
\end{equation}
Here we introduced the $\mathbb{Z}_3$-clock operators (qutrits) $X_{j,\mu},Z_{j,\mu}$ (for $\mu=A,B$) satisfying
$Z_{j,\mu}X_{j,\mu}=\omega X_{j,\mu}Z_{j,\mu}$ and
$X_{j,\mu}^3=Z_{j,\mu}^3=1$, where $\omega=e^{2\pi i/3}$.
Let us write $|\{a_j,b_j\}\rangle$ for the basis states such that
\begin{align}
  \begin{split}
    Z_{j,A}|\{a_j,b_j\}\rangle&=\omega^{a_j}|\{a_j,b_j\}\rangle,\\
    Z_{j,B}|\{a_j,b_j\}\rangle&=\omega^{b_j}|\{a_j,b_j\}\rangle.
  \end{split}
\end{align}
To define symmetry operators for $\mathbb{Z}_3^A\times S_3$, we consider the following operator:
\begin{equation}
  U_{\sigma}\coloneqq \prod_{j}C_{j,B}\sigma_{j}^x,
\end{equation}
where $C_{j,B}$ acts as $C_{j,B}|b\rangle_{j,B}=|-b\rangle_{j,B}$ in the $Z$-basis and $\sigma_{j}^x$ is a Pauli-$x$ operator acting on $\mathbb{C}^2$ for the $j$-th site.
To implement an embedding $\mathbb{Z}_3^A\times\mathbb{Z}_3^B \hookrightarrow \mathbb{Z}_3^A\times S_3$ twisted by a two cocycle $\beta\in \mathrm{H}^2(\mathbb{Z}_3^A\times\mathbb{Z}_3^B,\U(1))$, we use the disentangler $V$ for this SPT phase \cite{Chen:2011pg}.
The action of $V$ in the clock basis is given by
\begin{equation}
  V\lvert\{a_j,b_j\}\rangle
  \coloneqq\omega^{\sum_j b_j(a_j-a_{j+1})}
  \lvert\{a_j,b_j\}\rangle.
\end{equation}
Here $a_j,b_j\in\mathbb{Z}_3$ label the eigenvalues
$\omega^{a_j},\omega^{b_j}$ of $Z_{j,A},Z_{j,B}$, respectively. This operator $V$ commutes with $U_A$ and $U_B$.
Then we define the $\mathbb{Z}_3^A\times S_3$ symmetry generated by
\begin{equation}
  U_{A},\quad U_{r}\coloneqq U_B,\quad U_{s}\coloneqq VU_{\sigma}V^{\dagger}=V^2U_{\sigma}.
\end{equation}
The $\U(1)^2$ symmetry is generated by two charges $Q_A,Q_B$ defined by
\begin{equation}
  \begin{aligned}
    Q_A&\coloneqq\sum_j q_{j,A},\qquad
    Q_B\coloneqq\sum_j q_{j,B},\\
    q_{j,\mu}&\coloneqq 1-\sum_{k=1,2}
      \frac{X_{j,\mu}^{k}}{1-\omega^{-k}},\qquad \mu=A,B.
  \end{aligned}
\end{equation}
The local charges have eigenvalues $0,1,2$, so
$e^{2\pi iQ_A/3}=U_A$ and $e^{2\pi iQ_B/3}=U_B$,
realizing the fixed embedding of the common
$\mathbb{Z}_3^A\times\mathbb{Z}_3^B$ subgroup.

Now we define a family of Hamiltonians with the symmetry span. We first introduce the following operators:
\begin{equation}
  \begin{aligned}
    T_j^{(1)}&=X_{j,B},&
    V_j^{(1)}&=Z_{j-1,B}X_{j,A}Z_{j,B}^{\dagger},\\
    T_j^{(2)}&=Z_{j,A}X_{j,B}^{\dagger}Z_{j+1,A}^{\dagger},&
    V_j^{(2)}&=X_{j,A}.
  \end{aligned}
\end{equation}
Then we define the local Hamiltonian density $h_j\coloneqq\sum_{\alpha=1,2}h_j^{(\alpha)}$ as
\begin{align}
  \begin{split}
    h_j^{(\alpha)}&\coloneqq\frac{i}{1-\omega^{-1}}
      \Bigl[-\frac{T_j^{(\alpha)}+T_{j-1}^{(\alpha)}}{2}
      -V_j^{(\alpha)}\\
      &\qquad +(T_j^{(\alpha)}+T_j^{(\alpha)\dagger})V_j^{(\alpha)}\\
      &\qquad +V_j^{(\alpha)}
      (T_{j-1}^{(\alpha)}+T_{j-1}^{(\alpha)\dagger})\Bigr]
      +\mathrm{h.c.}
  \end{split}
\end{align}
Using this local Hamiltonian density, a family of Hamiltonians with the symmetry span is given by:
\begin{equation}
  H=H_{\sigma}+\sum_{n\geq 2}J_{n}H_{n},
\end{equation}
where $\{J_n\}_{n\geq 2}$ are real coupling constants, only finitely many of which are nonzero. Each term in the Hamiltonian is defined as
\begin{gather}
  H_{\sigma} = -\sum_{j}\sigma_j^x,\\
  H_2=\sum_{j}h_j,\quad B=\sum_{j}jh_{j},\quad H_{n+1}=i[H_n,B].
\end{gather}
We use this definition formally for the infinite chain, and for a sufficiently long finite periodic chain $H_n$ is defined by periodizing the resulting local density.
In this way, $H_3$ is given as $H_3=i\sum_{j}[h_j,h_{j+1}]$.
See \cite{Vernier:2018han} for an explicit construction of the underlying clock-chain charges on finite periodic chains using the transfer matrix.
With this prescription, each $H_n$, as well as $H_\sigma$, commutes with $Q_A$, $Q_B$, and $U_s$. Thus the Hamiltonian preserves the symmetry span for arbitrary values of the couplings.

To see the ground state of this Hamiltonian, let us consider the case where $J_2\neq0$ and $J_n=0$ for other $n$.
If one gauges the $\mathbb{Z}_3^A$ symmetry, the Hamiltonian becomes two decoupled copies of the self-dual $\U(1)$-invariant $\mathbb{Z}_3$ clock model studied in \cite{Vernier:2018han}, together with the gapped qubit sector described by $H_\sigma$.
Each copy flows to a conformal field theory (CFT) with central charge $c=3/2$ for $J_2>0$ and $c=1$ for $J_2<0$.
Since gauging a finite symmetry does not change the central charge, the original model flows to a CFT with $c=3$ for $J_2>0$ and $c=2$ for $J_2<0$.
The ground state and infrared theory of this model for more general values of the couplings have not been systematically studied in the literature.
However, the symmetry span argument excludes a gapped phase with finitely many ground states for any choice of the couplings.

\bigskip

\textit{Enforced gaplessness by finite symmetry}-----
Consider a symmetry span of the form
\begin{equation}
  \mathcal{C} \hookleftarrow \mathcal{E} \hookrightarrow \mathcal{D},
\end{equation}
where all symmetries are finite, namely some higher fusion categories.
If all symmetries are group symmetries, then no symmetry span can enforce gaplessness, because the fully broken phases of the two groups can be consistently pulled back to a fully broken phase of the shared group, with no unbroken symmetry and no freedom to stack distinct SPT phases.
Therefore, either of $\mathcal{C}$ or $\mathcal{D}$ has to be a non-invertible symmetry.

A simple example in 1+1D is
\begin{equation}
  \label{eq:finite-ising-dihedral-span}
  \TY(\mathbb{Z}_2,+)\hookleftarrow\Vect_{\mathbb{Z}_2}
  \hookrightarrow\Vect_{D_{10}}^{\omega},
\end{equation}
$\TY(\mathbb{Z}_2,+)$ is the Ising fusion category with positive Frobenius--Schur indicator, and $D_{10}=\langle r,s\mid r^5=s^2=1,\ srs=r^{-1}\rangle$ has order ten.
The common $\mathbb{Z}_2$ is generated by $s$, and $[\omega]$ is an order-five anomaly class, nontrivial on $\langle r\rangle$ and trivial on $\langle s\rangle$.
Let $N$ count $\mathbb{Z}_2$-fixed vacua and $M$ count exchanged pairs of vacua.
The Ising side requires $(N,M)=(k,k)$ \cite{Ando:2026ffy}.
On the group side, the only anomaly-free stabilizers are $1$ and the reflection subgroups, giving $(N,M)=(0,5)$ and $(1,2)$, respectively.
Thus any finite direct sum has $(N,M)=(m,2m+5n)$, which only has a solution to the Ising constraint if $m=n=0$.
The gaplessness conclusion is independent of the Frobenius--Schur sign, since $\TY(\mathbb{Z}_2,-)$ imposes the same vacuum-counting constraint $(N,M)=(k,k)$.

\medskip
A concrete example of a continuum theory realizing this symmetry span is given by a $c=3$ torus CFT, obtained by gauging a $\mathbb{Z}_2$ subgroup of the $\SU(2)_1^2\times\U(1)_{12}$ Wess--Zumino--Witten model.
This CFT is self-dual under gauging a non-anomalous $\mathbb{Z}_2$ symmetry $s$, up to T-duality, so that the gauging interface becomes a non-invertible duality defect $\mathcal{N}$.
Then $s$ and $\mathcal{N}$ generate a $\TY(\mathbb{Z}_2)$ symmetry.
The left-moving $\SU(2)$ symmetries of the parent theory survive the gauging, and an order-five rotation $r$ in them with $srs^{-1}=r^{-1}$ generates, together with $s$, a $D_{10}$ symmetry.
We can check that the $D_{10}$ symmetry has the anomaly $[\omega]$.
We provide the explicit construction of $\mathrm{TY}(\mathbb{Z}_2)$ symmetry and the $D_{10}$ symmetry in Appendix~\ref{app:torus-cft} of the End Matter, with further details in Supplemental Material, Sec.~\ref{sm:torus-cft}.

\medskip
Though the above example uses anomalous symmetries $\TY(\mathbb{Z}_2,+)$ and $\Vect_{D_{10}}^{\omega}$, 
we also expect that one can find a symmetry span with only non-anomalous finite symmetries that enforces gaplessness.
Indeed, OpenAI's ChatGPT (GPT-6 Astra) provided us with such an example together with a continuum CFT realization; we leave its detailed study for future work.
Hence we omit the details of this example in this letter.

\bigskip
\textit{Conclusion}-----
In this paper, we have shown that symmetry spans can enforce gaplessness using only non-anomalous invertible symmetries or only finite symmetries.
The gapped phases allowed by the two symmetries are incompatible upon restriction to the shared symmetry, even when spontaneous symmetry breaking is allowed.
We have also provided continuum realizations of both classes, together with a lattice realization of the former class.
It would be interesting to construct lattice realizations of the spans involving only finite symmetries and of symmetry spans in higher dimensions, with finite-dimensional local Hilbert spaces in both cases.

\bigskip
\textit{Acknowledgments}-----
We acknowledge substantial assistance from OpenAI's GPT-6 Astra in finding mathematical examples and constructing model realizations of the symmetry spans.
We also used it in drafting and revising the manuscript.

T.A.~is supported by JSPS KAKENHI Grant Number 25KJ1557.
K.O.~is supported by JST PRESTO Grant Number JPMJPR26A3, Japan, JSPS KAKENHI Grant-in-Aid No. 24K00522, and RIKEN ECL program.
K.O.~also acknowledges support by the Simons Foundation Grant \#888984 (Simons Collaboration on Global Categorical Symmetries).

\bibliography{ref}

\clearpage
\appendix
\renewcommand{\theHequation}{\thesection.\arabic{equation}}
{\centering\textbf{End Matter}\par}
\nopagebreak[4]
\medskip
\refstepcounter{section}\label{app:finite-group-classification}
\textit{Appendix \thesection: Gapped phases with finite group symmetry in 1+1D}---
Here we describe the general classification of SSB phases of finite
zero-form group symmetries $G$ in 1+1D, with anomaly $\omega\in\mathrm{H}^3(G,\U(1))$.
In an allowed SSB pattern, the unbroken symmetry $L$ has to be anomaly-free and one has a trivialization of the anomaly:
\begin{equation}
  \delta\alpha=\omega|_L,\quad \alpha\in \mathrm{C}^2(L,\U(1)).
\end{equation}
Then an allowed SSB pattern is specified by a subgroup $L\subset G$ that trivializes the anomaly.
Consider an indecomposable gapped phase, where the symmetry $G$ then acts transitively on the set of ground states.
For a chosen reference ground state $\Phi$ with unbroken symmetry $L$, we identify the vacuum set with the left coset space $G/L$:
\begin{equation}
  \Phi_{gL}\coloneqq g\cdot\Phi,\qquad
  k\cdot\Phi_{gL}=\Phi_{kgL}\quad(k\in G).
\end{equation}
There are $|G|/|L|$ vacua, and the unbroken symmetry in the sector $\Phi_{gL}$ is $L_g\coloneqq gLg^{-1}$.
The cochain $\alpha$ determines the anomaly trivialization in the reference ground state, while the ground state set and the permutation action of $G$ depend only on $L$.

After specifying the unbroken symmetry, we also need to specify an  SPT phase over the unbroken symmetry.
For a chosen reference ground state, this freedom is described by multiplying $\alpha$ by a two-cocycle $\eta\in\mathrm{Z}^2(L,\U(1))$.
The trivializations modulo coboundaries thus form an $\mathrm{H}^2(L,\U(1))$-torsor.
The data in the other ground state sectors are fixed by the $G$ symmetry.
Specifically, the anomaly trivialization $\alpha_g\in\mathrm{C}^2(L_g,\U(1))$ in the sector $\Phi_{gL}$ is given, up to a coboundary, by \cite{natale2017equivalence}
\begin{equation}
  \label{eq:sector-trivialization}
  \begin{gathered}
    \alpha_g=(c_{g^{-1}})^*\!\left(\frac{\alpha}{\Omega_g|_L}\right),\\
    \Omega_g(x,y)=\frac{\omega(c_g(x),c_g(y),g)\omega(g,x,y)}
    {\omega(c_g(x),g,y)},
  \end{gathered}
\end{equation}
where $c_g(x)=gxg^{-1}$
\footnote{Here $\omega$ denotes a normalized cocycle representative, satisfying $\omega(e,g,h)=\omega(g,e,h)=\omega(g,h,e)=1$ for the identity element $e\in G$. This can always be achieved by multiplying a cocycle representative by a coboundary, without changing its cohomology class. For any three-cocycle $\protect\widetilde{\omega}$, define $\beta(g,h)=\protect\widetilde{\omega}(g,e,e)/\protect\widetilde{\omega}(e,e,h)$. The cocycle condition implies that $\omega=\protect\widetilde{\omega}\,\delta\beta$ is normalized.}.
The identity $\delta\Omega_g=\omega/(c_g^*\omega)$ \cite[Eq.~(3.1)]{natale2017equivalence} ensures $\delta\alpha_g=\omega|_{L_g}$.

The relative SPT class $[\eta]$ is transported to $(c_{g^{-1}})^*[\eta]\in\mathrm{H}^2(L_g,\U(1))$.
In particular, when $\omega=1$, the sector $\Phi_{gL}$ carries the $L_g$ SPT phase
\begin{equation}
  [\alpha_g]=(c_{g^{-1}})^*[\alpha]
  \in\mathrm{H}^2(L_g,\U(1)).
\end{equation}
Thus the SPT phases in different sectors cannot be chosen independently; for non-anomalous abelian $G$, they are all the same $L$ SPT phase.

The pairs $(L,\alpha)$ and $(L_g,\alpha_g)$ describe the same gapped phase, with different choices of reference ground state.
Furthermore, multiplying $\alpha$ by a coboundary leaves the phase unchanged.
Therefore, for a fixed unbroken subgroup $L$, the indecomposable gapped phases are classified by the orbit set \cite[Theorem~1.1]{natale2017equivalence}:
\begin{equation}
  \label{eq:fixed-L-phases}
  \left\{[\alpha]\;\middle|\;\delta\alpha=\omega|_L\right\}
  \big/\bigl(N_G(L)/L\bigr).
\end{equation}
Here $[\alpha]$ is a trivialization modulo coboundaries, and $N_G(L)=\{g\in G\mid gLg^{-1}=L\}$ is the normalizer of $L$.
The coset $gL$ acts by $[\alpha]\mapsto[\alpha_g]$ according to \eqref{eq:sector-trivialization}.
Elements of $L$ act trivially on these classes, so the action descends to $N_G(L)/L$.
General gapped phases are direct sums of the above indecomposable phases.

\medskip
\refstepcounter{section}\label{app:torus-cft}
\textit{Appendix \thesection: Torus CFT realization of the finite span}---
A gapless theory with the symmetry span \eqref{eq:finite-ising-dihedral-span} is realized by a $c=3$ torus CFT, obtained by gauging a $\mathbb{Z}_2$ subgroup of the $\SU(2)_1^2\times\U(1)_{12}$ Wess--Zumino--Witten CFT.
We describe the resulting torus by its metric $\hat{G}$ and antisymmetric tensor $\hat{B}$~\cite{Giveon:1988tt}.
For angular coordinates of period $2\pi$, and integer momentum and winding charges $\hat{M},\hat{N}\in\mathbb{Z}^3$, the background is
\begin{equation}\label{eq:torus-gauged-background}
 \hat{E}=\hat{G}+\hat{B}=
 \begin{pmatrix}2&1&1\\0&1&0\\0&0&1\end{pmatrix}.
\end{equation}
Here $\hat{G}$ and $\hat{B}$ are the symmetric and antisymmetric parts of $\hat{E}$.
The change of charge basis and the first gauging are given in Supplemental Material, Sec.~\ref{sm:torus-cft}.

To check that this theory has $\mathrm{TY}(\mathbb{Z}_2)$ symmetry, we consider a $\mathbb{Z}_2$ symmetry given by
\begin{equation}\label{eq:torus-common-z2}
  sV_{\hat{M},\hat{N}}=(-1)^{\hat{M}_1}V_{\hat{M},\hat{N}}.
\end{equation}
The symmetry $s$ is a half-period shift of the first coordinate and is non-anomalous.
Gauging $s$ keeps even $\hat{M}_1$ and adds half-integer $\hat{N}_1$ sectors.
With $D=\operatorname{diag}(1/2,1,1)$, the background in an integer charge basis is
\begin{equation}\label{eq:torus-double-gauged-background}
  \check{E}=D\hat{E} D=
  \begin{pmatrix}
    1/2&1/2&1/2\\
    0&1&0\\
    0&0&1
  \end{pmatrix}.
\end{equation}
Denote the integer momentum and winding charges after this gauging by $\check{M},\check{N}$.
Now apply the factorized T-duality $T_1$ along the first coordinate~\cite{Buscher:1987sk,Buscher:1987qj,Giveon:1994fu}, which exchanges $\check{M}_1$ and $\check{N}_1$ and leaves the other charges unchanged.
On the background it acts by the $\mathrm{O}(3,3;\mathbb{Z})$ transformation $E\mapsto(aE+b)(cE+d)^{-1}$ with $a=d=\operatorname{diag}(0,1,1)$ and $b=c=\operatorname{diag}(1,0,0)$~\cite{Giveon:1988tt}, which transforms the background to
\begin{equation}
  T_1(\check{E})=
  \begin{pmatrix}
    2&-1&-1\\
    0&1&0\\
    0&0&1
  \end{pmatrix}.
\end{equation}
A subsequent inversion of the first coordinate, represented by $S=\operatorname{diag}(-1,1,1)$, gives
$S^T T_1(\check{E})S=\hat{E}$.
Since T-duality gives an equivalent CFT~\cite{Rocek:1991ps}, the theory with background $\hat{E}$ is self-dual under gauging $s$.
The combined identification sends $(\check{M}_1,\check{N}_1)$ to $(-\check{N}_1,-\check{M}_1)$, leaving the other charges unchanged.
In particular, the quantum $\mathbb{Z}_2$ symmetry $(-1)^{\check{N}_1}$ of this second gauging is identified with $s$.

The gauging interface therefore becomes a non-invertible duality defect $\mathcal{N}$, with $\mathcal{N}^2=1+s$ and $s\mathcal{N}=\mathcal{N} s=\mathcal{N}$.
Thus $s$ and $\mathcal{N}$ generate a $\TY(\mathbb{Z}_2)$ symmetry.

We now construct a $D_{10}$ symmetry sharing the same $s$.
We introduce the charges
\begin{equation}\label{eq:torus-left-charges}
  Q_1=\hat{M}_2+\hat{N}_1+\hat{N}_2,\quad
 Q_2=\hat{M}_3+\hat{N}_1+\hat{N}_3.
\end{equation}
In terms of the original momentum and winding charges $m,n$, they are given by $Q_i=m_i+n_i$ for $i=1,2$.
Thus $Q_i/2$ is the Cartan charge of the left-moving $\SU(2)$ symmetry of the $i$th original $\SU(2)_1$ factor.
Since these two $\SU(2)$ actions commute with the gauged $\mathbb{Z}_2$ symmetry, they remain symmetries after gauging.
We first consider two operators $r_1,s_1$ acting as
\begin{equation}\label{eq:torus-auxiliary-dihedral}
  \begin{gathered}
  \begin{aligned}
    r_1V_{\hat{M},\hat{N}}&=e^{\frac{2\pi i}{5}(Q_1+Q_2)}V_{\hat{M},\hat{N}},\\
    s_1V_{\hat{M},\hat{N}}&=(-1)^{\hat{M}_1+\hat{M}_2Q_1+\hat{M}_3Q_2}
    V_{\hat{M}',\hat{N}'},
  \end{aligned}\\
  \begin{aligned}
    \hat{M}'_1&=\hat{M}_1-Q_1-Q_2,&\quad \hat{N}'_1&=\hat{N}_1,\\
    \hat{M}'_2&=\hat{M}_2-Q_1,&\quad \hat{N}'_2&=\hat{N}_2-Q_1,\\
    \hat{M}'_3&=\hat{M}_3-Q_2,&\quad \hat{N}'_3&=\hat{N}_3-Q_2.
  \end{aligned}
  \end{gathered}
\end{equation}
One can check that this action generates a $D_{10}$ symmetry by verifying that it is invertible on the integer charge lattice, preserves $h,\bar{h}$ at fixed $\hat{E}$ and the vertex-operator OPE phases, and satisfies the group relations.

To relate $s_1$ to the TY generator $s$, we use rotations of the two original left-moving $\SU(2)$ factors, which survive the first gauging.
With suitable phase conventions, we take $U_i$ to be the rotation in the $i$th factor with fundamental-representation matrix $u$, where
\begin{equation}\label{eq:torus-conjugating-rotation}
  u=\frac1{\sqrt2}
  \begin{pmatrix}1&1\\-1&1\end{pmatrix},
  \qquad U=U_1U_2.
\end{equation}
Here $U_1,U_2$ act in different factors and commute.
As exponentials of the zero modes of weight-$(1,0)$ currents, these rotations preserve both Virasoro algebras and the OPE~\cite{HarveyMoore:2018}.
Thus $U$ is an internal symmetry of the gauged CFT at fixed $\hat{G},\hat{B}$.
One can check on the vertex operators generating the charge lattice that $U^{-1}s_1U$ acts by $(-1)^{\hat{M}_1}$ without changing the charges.
Since both actions preserve the OPE, their agreement on these generators implies $s=U^{-1}s_1U$.
Let us define $r\coloneqq U^{-1}r_1U$, then the group relations of $r_1,s_1$ imply $r^5=s^2=1,srs^{-1}=r^{-1}$.
Since $U$ and $r_1$ are CFT symmetries, $r$ is also a CFT symmetry.
Therefore, $r,s$ generate a $D_{10}$ symmetry whose reflection is the TY generator $s$ in \eqref{eq:torus-common-z2}.

Finally, we compute the anomaly of this $D_{10}$ symmetry using the conjugate generators $r_1,s_1$.
The phase of $r_1$ in \eqref{eq:torus-auxiliary-dihedral} is determined by
\begin{equation}
  Q_1+Q_2=\hat{M}_2+\hat{M}_3+2\hat{N}_1+\hat{N}_2+\hat{N}_3.
\end{equation}
For an action of the form $e^{2\pi i(\cdots)/5}$, the $\mathbb{Z}_5$ anomaly is given by the sum of the products of the $\hat{M}_i$ and $\hat{N}_i$ coefficients, modulo five~\cite[Sec.~3.2]{LinShao:2021}.
The anomaly of $r_1$ is therefore $0\times2+1\times1+1\times1=2\pmod5$.
Since $s_1$ is conjugate to the non-anomalous half-period shift $s$, its $\mathbb{Z}_2$ anomaly is trivial.
These restrictions determine an order-five class in $\mathrm{H}^3(D_{10},\U(1))\cong\mathbb{Z}_{10}$~\cite[Theorem~5.3]{Handel:1993}.
This gives the anomaly $[\omega]$ required in \eqref{eq:finite-ising-dihedral-span}.

\medskip
\refstepcounter{section}\label{app:higher-dimensional}
\textit{Appendix \thesection: A symmetry span in 2+1D}-----
Here we give an example of a symmetry span in 2+1D that enforces gaplessness even when intrinsic topological order is allowed, using only non-anomalous invertible symmetries.
Consider a bosonic span
\begin{equation}\label{eq:higher-u1-span}
  S_3\times\mathbb{Z}_3^2
  \xhookleftarrow{\;i_{\beta}\;}\mathbb{Z}_3^3
  \xhookrightarrow{\;j\;}\U(1)^3,
\end{equation}
Here the first factor of $\mathbb{Z}_3^3$ embeds into the three-cycle subgroup of $S_3$, while the other two embed into the direct factors, and $j$ is the standard inclusion.
$\beta\in\mathrm{H}^3(\mathbb{Z}_3^3,\U(1))$ is the type-III SPT class, whose topological response phase in 3D spacetime is $\beta(A)=\exp(\frac{2\pi i}{3}\int_M A_a A_b A_c)$,
where $A_a,A_b,A_c$ are $\mathbb{Z}_3$ backgrounds on a closed oriented three-manifold $M$.
Let us write $H=\mathbb{Z}_3^3$, $G=S_3\times\mathbb{Z}_3^2$, and $K=\U(1)^3$.
Since we twist the embedding into $G$ by the type-III $H$-SPT, compatibility requires that the partition function $Z$ satisfies
\begin{equation}\label{eq:higher-span-matching}
  \beta(A)\,Z[i(A)]=Z[j(A)].
\end{equation}

Let us show this requirement never holds for any gapped phase with finitely many ground states.
To see this, we first note that the order-two reflection in $S_3$ induces $\sigma(a,b,c)=(-a,b,c)$ and sends $\beta$ to $\beta^{-1}$.
Since $i(A)$ and $i(\sigma A)$ are gauge-equivalent $G$ backgrounds, the non-anomalous $G$ theory satisfies
\begin{equation}\label{eq:higher-conjugacy}
  Z[i(\sigma A)]=Z[i(A)].
\end{equation}
This is gauge invariance of the full theory, not an assumption that each vacuum preserves the reflection.

Now place the $a,b,c$ holonomies on the three separate cycles of $T^3$.
Since $H^2(T^3,\mathbb{Z})$ has no torsion, the induced flat $\U(1)$ bundles are topologically trivial, so both $\U(1)^3$ backgrounds admit globally defined flat connections.
Since $K=\U(1)^3$ is connected, it cannot permute the finitely many vacua or anyon types of a finite TQFT.
Although $\U(1)^3$ fractionalization may be nontrivial, its coupling depends on the first Chern classes, which vanish on these backgrounds.
Moreover, the background Chern--Simons response of any $\U(1)^3$-SPT is also trivial on these globally defined flat connections.
Consequently,
\begin{equation}\label{eq:higher-torus-positive}
  Z[j(A)]=Z[j(\sigma A)]=D,
  \quad D=\dim\mathcal{H}(T^2)>0.
\end{equation}
Since $\beta(A)=\omega:=e^{2\pi i/3}$ and $\beta(\sigma A)=\omega^{-1}$, \eqref{eq:higher-span-matching} and \eqref{eq:higher-conjugacy} would require $\omega^{-1}D=\omega D$, but this is a contradiction.
Allowing SSB and reducible TQFTs does not evade this argument: $D$ is then a sum of positive state-space dimensions over all vacua.
For example, if $S_3\ssb\mathbb{Z}_3$ gives two exchanged SET sectors $\pm$, their dimensions would have to satisfy $\omega^{-1}D_+=\omega D_-$ with $D_\pm>0$, which is also impossible.
Continuum field theories realizing this span are presented in Sec.~\ref{sm:higher-dimensional} of the Supplemental Material.

\clearpage
\onecolumngrid
\setcounter{section}{0}
\setcounter{subsection}{0}
\setcounter{equation}{0}
\renewcommand{\thesection}{\Roman{section}}
\renewcommand{\thesubsection}{\thesection.\arabic{subsection}}
\renewcommand{\theHsection}{SM.\arabic{section}}
\renewcommand{\theHsubsection}{SM.\arabic{section}.\arabic{subsection}}
\renewcommand{\theequation}{S\arabic{equation}}
\renewcommand{\theHequation}{SM.\arabic{equation}}
\makeatletter
\def\p@subsection{}
\@removefromreset{equation}{section}
\let\@sectioncntformat\@seccntformat
\def\@hangfrom@section#1#2#3{\@hangfrom{#1#2}#3}
\makeatother
\allowdisplaybreaks
{\centering\textbf{Supplemental Material}\par}

\section{Invertible symmetry spans and their continuum realizations}\label{sm:subgroup-data}\label{sm:invertible-continuum}
Here we provide another example of a symmetry span that enforces gaplessness using only non-anomalous invertible symmetries, together with a continuum realization.
The span is given by
\begin{equation}\label{eq:span_invertible_D8}
  \U(1)^3\hookleftarrow\mathbb{Z}_2^3\xhookrightarrow{\beta^\prime}D_8\times\mathbb{Z}_2,
\end{equation}
where $\beta^\prime\in\mathrm{H}^2(\mathbb{Z}_2^3,\U(1))$ is a nontrivial SPT class specified below.
We write $G=D_8\times\mathbb{Z}_2$, where $D_8=\langle r,s\mid r^4=s^2=1,\ srs=r^{-1}\rangle$ has order eight and $z$ generates the extra $\mathbb{Z}_2$.
The shared subgroup is $H=\langle x,y,z\rangle\cong\mathbb{Z}_2^3$, with $x=s$ and $y=r^2s$.

Consider the SSB pattern $G\ssb L$ with $L=H$.
We label the SPT classes in $\mathrm{H}^2(L,\U(1))\cong\mathbb{Z}_2^3$ by $(p,q,t)$, with cocycle representatives
\begin{equation}
  \alpha_{p,q,t}\bigl(x^a y^b z^c,\ x^{a'}y^{b'}z^{c'}\bigr)
  =(-1)^{pab'+qac'+tbc'},
\end{equation}
where all exponents take values in $\mathbb{Z}_2$.
The three classes $[\alpha_{1,0,0}]$, $[\alpha_{0,1,0}]$, and $[\alpha_{0,0,1}]$ generate $\mathrm{H}^2(L,\U(1))$.
Conjugation by $r$ exchanges $x$ and $y$ and fixes $z$, sending $[\alpha_{p,q,t}]$ to $[\alpha_{p,t,q}]$.
Thus, for an indecomposable gapped phase with unbroken subgroup $L$, the SPT phases in the two ground state sectors are labeled by $(p,q,t)$ and $(p,t,q)$.
There are therefore six indecomposable gapped phases with this unbroken subgroup: four corresponding to the fixed points with $q=t$ and two corresponding to the two-element orbits with $q\ne t$.
We choose $\beta^\prime=[\alpha_{p,q,t}]$ with $q\ne t$.

We now compare the pullbacks of gapped phases through the two embeddings.
Because of the left embedding, the shared $H$ symmetry must be unbroken in any gapped phase, and the pullback from $\U(1)^3$ gives only the trivial $H$-SPT phase and its direct sums.
Since $H$ must remain unbroken, $G$ is either spontaneously broken to $H$, with $L=H$, or remains unbroken, with $L=G$.
For $L=H$, both ground state sectors would have to carry the $H$-SPT class $(\beta^\prime)^{-1}$ before the twist.
This is impossible because $r$ exchanges the two ground state sectors but does not preserve the class $(\beta^\prime)^{-1}$.
For $L=G$, the restriction of a $G$-SPT class to $H$ is invariant under conjugation by $r$, so it cannot equal $(\beta^\prime)^{-1}$ either.
Thus no gapped phase is compatible with both embeddings, and the symmetry span enforces gaplessness.

We next construct a gapless theory realizing this span.
Consider three decoupled compact bosons $\phi_i\sim\phi_i+2\pi$, with $c=3$, and choose the first two radii to be equal.
We denote their integer momentum and winding charges by $P_i$ and $W_i$, respectively.
The generators $x,y,z$ act as half-period shifts of $\phi_1,\phi_2,\phi_3$, respectively.
Together with the exchange of the first two bosons, these shifts generate $D_8\times\mathbb{Z}_2$.
We take the $\U(1)^3$ charges to be
\begin{equation}
  Q_1=P_1+2W_3,\quad Q_2=P_2,\quad Q_3=P_3-2W_1.
\end{equation}
Since $e^{i\pi Q_i}=e^{i\pi P_i}$, the two symmetries share the same $\mathbb{Z}_2^3$ subgroup.
The relative SPT class of the two embeddings is $\beta^\prime=[\alpha_{0,1,0}]$\footnote{This can be checked by restricting the background-field couplings of the two symmetry groups to the common $\mathbb{Z}_2^3$ subgroup and computing the ratio of the resulting partition functions.}.
The $\U(1)^3$ symmetry is non-anomalous because the mixed anomalies between the shift and winding symmetries cancel each other.
The $D_8\times\mathbb{Z}_2$ symmetry acts geometrically and is also non-anomalous.
In this realization, conjugation by $r$ maps the $\U(1)^3$ symmetry to a different $\U(1)^3$ with charges $P_2+2W_3$, $P_1$, and $P_3-2W_2$.
These two $\U(1)^3$ symmetries generate a $\U(1)^5$ symmetry, whose mixed anomaly by itself forbids a gapped phase.

\section{Details of the torus CFT realization}\label{sm:torus-cft}
Here we give the charge-basis transformations and the gauging of the parent theory used in Appendix~\ref{app:torus-cft} of the End Matter.

Let us start with the $\SU(2)^2_1\times\U(1)_{12}$ Wess--Zumino--Witten CFT, realized by three compact bosons with radii $(1,1,\sqrt{6})$ in units where the self-dual radius is one.
We describe the torus by its metric and antisymmetric tensor~\cite{Giveon:1988tt}.
For angular coordinates of period $2\pi$, the background metric $G$ and the $B$-field are
\begin{equation*}
  G=\operatorname{diag}(1,1,6),\qquad B=0.
\end{equation*}
Let $m,n\in\mathbb{Z}^3$ denote the integer momentum and winding charges of the vertex operators $V_{m,n}$.
The $\mathbb{Z}_2$ symmetry to be gauged acts as
\begin{equation}
  gV_{m,n}=(-1)^{m_1+n_1+m_2+n_2+m_3}V_{m,n}.
\end{equation}
After changing the basis of the momentum--winding charge lattice to $M_i$ and $N_i$ defined by
\begin{equation}
  \begin{aligned}
    M_1&=m_1+m_2+m_3+n_1+n_2-2n_3,
          &N_1&=n_3,\\
    M_2&=m_1-n_3,
          &N_2&=n_1-n_3,\\
    M_3&=m_2-n_3,
          &N_3&=n_2-n_3,
  \end{aligned}
\end{equation}
then the $\mathbb{Z}_2$ action is given by $(-1)^{M_1}V_{M,N}$.
In this basis, the background matrix is
\begin{equation}
  E^\prime=G^\prime+B^\prime=
  \begin{pmatrix}
    8 &  2 & 2\\
    0 & 1 & 0\\
    0 & 0 & 1
  \end{pmatrix}.
\end{equation}
Taking its symmetric and antisymmetric parts gives
\begin{equation}
  G^\prime=
  \begin{pmatrix}
    8 & 1 & 1\\
    1 & 1 & 0\\
    1 & 0 & 1
  \end{pmatrix},\quad
  B^\prime=
  \begin{pmatrix}
    0 & 1 & 1\\
    -1 & 0 & 0\\
    -1 & 0 & 0
  \end{pmatrix}.
\end{equation}
Now gauge $g$, which is a half-period shift in the new basis.
The projection keeps states with even $M_1$, while the twisted sector adds half-integer $N_1$.
We therefore introduce integer momentum and winding charges $\hat{M},\hat{N}\in\mathbb{Z}^3$ by $M=D^{-1}\hat{M}$ and $N=D\hat{N}$, where $D=\mathrm{diag}(1/2,1,1)$.
In this charge basis, the background transforms as
\begin{equation}\label{eq:sm-torus-background-transform}
 \hat{G}=DG^\prime D,\qquad \hat{B}=DB^\prime D,\qquad
 \hat{E}=DE^\prime D,
\end{equation}
which gives \eqref{eq:torus-gauged-background}.

\section{Continuum realizations of the 2+1D symmetry span}\label{sm:higher-dimensional}
Here we construct a family of continuum theories realizing the symmetry span \eqref{eq:higher-u1-span} in Appendix~\ref{app:higher-dimensional} of the End Matter.
We use the same notation $H=\mathbb{Z}_3^3$, $G=S_3\times\mathbb{Z}_3^2$, and $K=\U(1)^3$.
Consider three compact scalar fields $\phi_i\sim\phi_i+2\pi$, with $i=a,b,c$, on a closed oriented three-dimensional spacetime manifold $M$.
The compact scalars are defined to be smooth circle-valued fields $e^{i\phi_i}\colon M\to\U(1)$, and integrated including all winding sectors.
Their Euclidean action is $S=S_{\mathrm{kin}}+S_{\mathrm{top}}$ with
\begin{equation}\label{eq:higher-continuum-top}
 S_{\mathrm{top}}=2\pi i\times 9\int_M
 \frac{d\phi_a}{2\pi}\wedge\frac{d\phi_b}{2\pi}\wedge\frac{d\phi_c}{2\pi}.
\end{equation}
Note that this topological term takes values in $2\pi i\mathbb{Z}$ on $M$, and so it does not affect the dynamics of the theory.
We take $S_{\mathrm{kin}}$ to be invariant under constant shifts and the reflection $\mathcal{S}\colon\phi_a\mapsto-\phi_a$.
The $K$ symmetry acts by constant shifts of the three fields, and its subgroup $H$ is generated by $g_i:\phi_i\mapsto\phi_i+2\pi/3$.
On a closed spatial surface $\Sigma$, let $W_{abc}\in\mathbb{R}/\mathbb{Z}$ denote the surface holonomy, which is locally written as $\int_\Sigma\frac{\phi_a}{2\pi}\frac{d\phi_b}{2\pi}\wedge\frac{d\phi_c}{2\pi}$.
Define the diagonal unitary $V$ and the reflection generator $R$ of $G$ by
\begin{equation}\label{eq:higher-continuum-reflection}
 V=e^{-2\pi i9W_{abc}},\qquad
 R=V\mathcal{S} V^{-1}.
\end{equation}
Note that $d(\phi_a\,d\phi_b\wedge d\phi_c)=d\phi_a\wedge d\phi_b\wedge d\phi_c$ locally.
Then one can show that
\begin{equation}\label{eq:higher-continuum-endpoint}
 e^{-S_{\mathrm{top}}}
 =V[\phi_{\mathrm{out}}]\,V[\phi_{\mathrm{in}}]^{-1}
\end{equation}
on a Euclidean time slab $\Sigma\times[0,T]$, with initial and final field configurations $\phi_{\mathrm{in}}$ and $\phi_{\mathrm{out}}$.
This identity holds even if $W_{abc}$ is defined only modulo integers~\cite{BarBecker:2014}.
Put differently, let $\widehat{H}_0$ and $\widehat{H}$ be the Hamiltonians for $S_{\mathrm{kin}}$ and $S_{\mathrm{kin}}+S_{\mathrm{top}}$, respectively.
The path integral with fixed initial and final fields gives a matrix element of the Euclidean evolution operator.
Since $V$ is diagonal in this field basis, the endpoint phases imply
\begin{equation}\label{eq:higher-continuum-conjugation}
 e^{-T\widehat{H}}=V e^{-T\widehat{H}_0}V^{-1},\qquad
 \widehat{H}=V\widehat{H}_0V^{-1},
\end{equation}
where the second equality follows by differentiating at $T=0$.
Thus the symmetry $\mathcal{S}$ of $\widehat{H}_0$ becomes the symmetry $R=V\mathcal{S} V^{-1}$ of $\widehat{H}$.
Since $V$ commutes with $H$, the operators obey $R^2=1$, $Rg_aR^{-1}=g_a^{-1}$, and $[R,g_b]=[R,g_c]=0$.
Thus $R,g_a,g_b,g_c$ generate $G$.
On the other hand, the $K$ symmetry couples to ordinary background gauge fields.

To compare the two embeddings of $H$, let $A_a,A_b,A_c$ be $\mathbb{Z}_3$ backgrounds on $M$, as in Appendix~\ref{app:higher-dimensional}.
First suppose that the induced $\U(1)^3$ background admits globally defined flat connections $\mathcal{A}_i$, with $\oint_\gamma\mathcal{A}_i=\frac{2\pi}{3}\oint_\gamma A_i\pmod{2\pi}$ for every closed curve $\gamma$.
Then the covariant derivatives are $D\phi_i=d\phi_i-\mathcal{A}_i$, and one can see that the two couplings differ only by $e^{-S_{\mathrm{top}}[D\phi]}$.
The field-dependent terms contribute integer multiples of $2\pi i$ to the action, while the remaining background phase is $\beta(A)$ defined in Appendix~\ref{app:higher-dimensional}.
Therefore,
\begin{equation}\label{eq:higher-continuum-matching}
 \beta(A)\,Z[i(A)]=Z[j(A)],
\end{equation}
which is the condition \eqref{eq:higher-span-matching}, independently of the choice of invariant kinetic term.
Note that, for a general background $A_i$, $e^{i\phi_i}$ is a section of the associated circle bundle.
Indeed, invariance of $D\phi_i$ under $\mathcal{A}_i\mapsto\mathcal{A}_i+d\lambda_i$ requires $e^{i\phi_i}\mapsto e^{i\lambda_i}e^{i\phi_i}$, which is the gluing law for a section on overlapping patches.
If this bundle is topologically nontrivial, it has no global section.
Both partition functions therefore vanish because there are no field configurations to integrate over.
We first choose a relativistic kinetic term and then consider deformations that preserve the symmetry span \eqref{eq:higher-u1-span}.

\medskip
\textit{Relativistic Goldstone phase}-----
We take
\begin{equation}\label{eq:higher-continuum-relativistic}
 S_{\mathrm{kin}}=\frac{f^2}{2}\sum_i
 \int_M d\phi_i\wedge *d\phi_i,\qquad f^2>0.
\end{equation}
This theory has three relativistic Goldstone modes with dispersion $\omega_{\mathbf{k}}=|\mathbf{k}|$ in units with velocity one.
The continuous $K$ symmetry is spontaneously broken in the ground state in infinite volume.
Indeed, a Gaussian calculation on $\mathbb{R}^3$ gives the large-distance behavior
\begin{equation}
 \bigl\langle e^{i\phi_i(x)}e^{-i\phi_i(0)}\bigr\rangle
 \simeq C_i\exp\left(\frac{1}{4\pi f^2|x|}\right)
 \longrightarrow C_i>0,
\end{equation}
where $C_i$ depends on the ultraviolet regularization and normalization of the vertex operators.

\medskip
\textit{Deformations preserving the symmetry span}-----
Starting from the relativistic theory \eqref{eq:higher-continuum-relativistic}, we can vary the coefficients and add spatial derivative terms to obtain
\begin{equation}\label{eq:higher-continuum-lifshitz}
 S_{\mathrm{kin}}=\int d\tau\,d^2x\sum_i\left[
 \frac{\chi}{2}(\partial_\tau\phi_i)^2
 +\frac{\kappa}{2}(\nabla^2\phi_i)^2
 +\frac{\rho}{2}|\nabla\phi_i|^2
 +\frac{\lambda}{4}|\nabla\phi_i|^4\right].
\end{equation}
These terms preserve $K$ and $\mathcal{S}$.
Keeping $S_{\mathrm{top}}$ and $V$ fixed therefore preserves $R=V\mathcal{S} V^{-1}$ and the same symmetry span, although Lorentz invariance is lost.

For $\chi,\kappa>0$, the Gaussian Lifshitz point $\rho=\lambda=0$ has infinitely many degenerate ground states on a spatial torus, labeled by winding numbers: constant gradients satisfy $\nabla^2\phi_i=0$ and cost no energy~\cite{Gorantla:2023lifshitz}.
With $\rho$ tuned so that the renormalized coefficient of $|\nabla\phi_i|^2$ vanishes, one-loop renormalization-group analyses find that a small positive $\lambda$ is marginally irrelevant~\cite{Vishwanath:2003yjl,Hsu:2013lifshitz}.
The resulting flow approaches the Gaussian $z=2$ fixed point with logarithmic corrections~\cite{Vishwanath:2003yjl}.
The entire family is consistent with the constraint imposed by the symmetry span.
The span excludes gapped phases described by TQFTs with finitely many ground states, but does not fix the dynamical exponent or the ground-state degeneracy.

This family also has a mixed anomaly between $K$ and its exact $\U(1)^3$ winding one-form symmetry, with normalized anomaly polynomial $I_5=(2\pi)^{-2}\sum_{i=a,b,c}d\mathcal{A}_i\wedge d\mathcal{B}_i$, where $\mathcal{A}_i$ and $\mathcal{B}_i$ are the respective one-form and two-form background gauge fields.
This anomaly does not admit any symmetry-preserving TQFT~\cite{Delacretaz:2019brr,Cordova:2024torsion}, and so gaplessness for these models actually follows from the mixed anomaly.

\end{document}